\documentclass[prl,twocolumn,superscriptaddress,showpacs,preprintnumbers,amsmath,amssymb]{revtex4}
\usepackage{amsmath}
\usepackage{graphicx}
\usepackage{dcolumn}

\begin{document}
\title {Interfacial confinement in core-shell nanowires due to high dielectric mismatch}

\author{A. A. Sousa}
\author{T. A. S. Pereira}
\affiliation{Institute of Physics, Federal University of Mato Grosso, 78060-900, Cuiab\'a, Mato Grosso, Brazil}

\author{A. Chaves}
\author{J. S. de Sousa}
\author{G. A. Farias}
\affiliation{Department of Physics, Federal University of Cear\'a, C.P. 6030, 60455-900, Fortaleza, Cear\'a, Brazil}


\begin{abstract}

We theoretically investigate the role of the dielectric mismatch
between materials on the energy levels and recombination energies
of a core-shell nanowire. Our results demonstrate that when the
dielectric constant of the core material is lower than that of the
shell material, the self-image potential pushes the charge
carriers towards the core-shell interface, in such a way that the
ideal confinement model is no longer suitable. The effects of this
interfacial confinement on the electronic properties of such
wires, as well as on its response to applied magnetic fields, are
discussed.

\end{abstract}

\maketitle

Great attention has been devoted to the
investigation of the electronics and optical properties of
core-shell nanowires (NW). In particular, the applications of
these low dimensionality structure in optoelectronic and photonic
devices are of interest for the electronics industry, and
much effort has been dedicated to their fabrication. \cite{mohan,persson,chen,tsivion} In addition, studies of low
dimensional systems surrounded by high dielectric constant
materials continue to attract attention from many researchers
\cite{Jena, zhu} towards a continuation of the Moore's law. Recently,
wire diameters of a few nanometers were experimentally achieved, \cite{Jiang} and carrier confinement effects in these nanowires
have been reported with different levels of sophistication. \cite{li, silva, slachmuylders}

This work aims to investigate the dielectric mismatch effects on
core-shell NW, focusing on the possibility of interfacial
confinement of the carriers. As for the model structure, we
consider a semiconductor cylindrical nanowire (core region) of
radius $R$, surrounded by a different material (shell region). The
interface between core and shell regions is assumed to be abrupt,
i.e. the materials parameters change abruptly from the core to the
shell regions. For the heterostructure materials considered in this paper, the bands mismatch creates a high potential barrier for the charge carriers in the shell, leading to a short penetration of the wave functions in this region. This rules out the role of the shell width on the energy states of the NW, since the wave functions does not reach the outer edge of the shell. The nanowire electronic structure is obtained by
solving numerically a Schr\"odinger-like equation within the
adiabatic approach and the effective mass framework.  \cite{Supplementary} The total
confinement potential $V_T^i(\rho_i) = \Delta
E_i(\rho_i)+\Sigma_i(\rho_i)$ is given by the sum of band edges
discontinuities $\Delta E_i(\rho_i)$ and the self-energy potential
$\Sigma_i(\rho_i)$, where $i=e,lh,hh$ represents the carrier
types (electron, light hole and heavy hole, respectively). The latter term appears due to the dielectric mismatch, and its
calculation is based on the method of the image charges. The
details of this calculation can be seen in our supplementary material. \cite{Supplementary} In a nutshell, the self-energy potential $\Sigma_i(\rho_i)$ inside the core region,
due to a carrier located in the core region, is given by 
Eq.(19) of Ref. \onlinecite{slachmuylders}, whereas when the
carrier is located in the shell region, the self-energy potential
inside this region can be obtained using a similar expression,
just by changing the modified Bessel functions of the first type,
$I_m^2(k\rho_e)$ and $I_m^2(k\rho_h)$, by those of the second
type, $K_m^2(k\rho_e)$ and $K_m^2(k\rho_h)$, respectively.

\begin{figure}[t]
\begin{center}
\includegraphics[width= 1.0\linewidth]{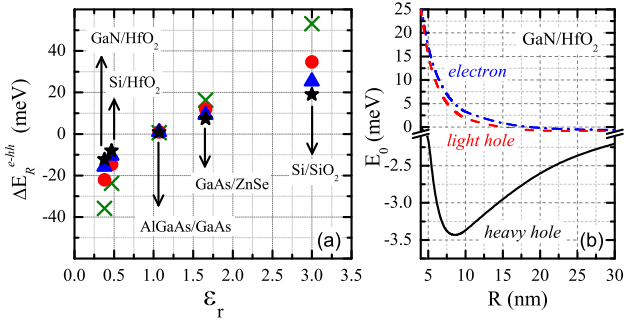}
\caption{\small{\label{fig1} (a) Effect of the self-energy
potential on the recombination energy of an electron-hole pair,
for different wire radii: 4 nm ($\times$), 6 nm ($\bullet$), 8 nm
($\blacktriangle$) and 10 nm ($\blacksquare$). (b) Ground state
energies as a function of the core radius for carriers confined in a
GaN/HfO$_2$ core-shell heterostructure.}}
\end{center}
\end{figure}

The electron-hole recombination energy, $E_R^{e-h} = E_G + E_e +
E_h$ ($h = lh$ or $hh$), given by the sum of the band gap energy
$E_G$ on the core region with the electron $E_e$ and hole $E_h$
confinement energies, is calculated for the radial ground state
quantum number $n=1$ and zero angular momentum $l=0$, allowing us
to analyze the effects caused by the self-energy potential on the
ground state energy of the electron-hole pair. Figure
\ref{fig1}(a) shows the role of the dielectric mismatch on the
value of the electron-hole recombination energy $\Delta E_R^{e-h}
= E_{R,\Sigma_i(\rho_i) \neq 0}-E_{R,\Sigma_i(\rho_i) = 0}$, as a
function of the ratio $\epsilon_r = \varepsilon_1 /
\varepsilon_2$ between dielectric constants of the core ($\varepsilon_1$) and shell ($\varepsilon_2$) materials, considering several values of nanowire radii $R$, for different materials. This quantity denotes the difference
between the recombination energies with ($E_{R,\Sigma_i(\rho_i)
\neq 0}$) and without ($E_{R,\Sigma_i(\rho_i) = 0}$) the
self-energy corrections. The analysis of this difference gives us
the advantage of excluding the effect of quantum confinement due
to the band edge mismatch $\Delta E_i$, remaining exclusively the
confinement due to the dielectric mismatch. The material
parameters are the same as in Ref. \onlinecite{pereira}.

As in the case of quantum wells, shown by Pereira \emph{et al.}
\cite{pereira}, if the dielectric constant in core region
$\varepsilon_1$ is smaller than that of the shell
region $\varepsilon_2$, the potential in the core region is
attractive and its contribution to the recombination energy is
negative $\Delta E_R < 0$. On the other hand, for a larger
dielectric constant in core region, as compared to the one in the
shell region, the potential in the core region is repulsive, so
that its contribution to the recombination energy is positive
$\Delta E_R > 0$, as clearly shown in Fig. \ref{fig1}(a). The
resulting $\Delta E_R^{e-h}$ for both light and heavy holes are
the same, as this quantity only contains the effects of
$\Sigma_i(\rho_i)$ on the carrier charge. On the other hand, by
plotting $E_R^{e-h}$, it is possible to observe the difference
between $e-lh$ and $e-hh$ pairs, caused by the different
light hole and heavy hole effective masses.

\begin{figure}[t]
\begin{center}
\includegraphics[width= 1.0\linewidth]{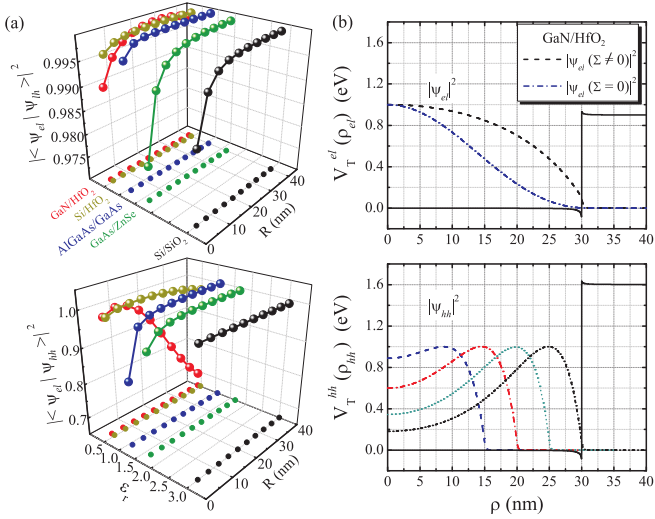}
\caption{\small{\label{fig2} (a) Overlap between the electron (\textit{el}) and the light (\textit{lh}) and heavy (\textit{hh}) hole wave functions as a function of the core radius $R$, for several heterostructure materials with different ratios between dielectric constants $\epsilon_r$, represented by different colors. (b) Effective confinement potentials $V_T^i$ (solid) for $R = 30$ nm in the GaN/HfO$_2$ case for electrons and heavy holes, along with some examples of their wave functions. Electron wave functions are shown with (dashed) and without (dashed-dotted) taking image charge effects into account, for comparison. Hole wave functions (for $\Sigma \neq 0$) are depicted for $R = $ 15 nm (blue dashed), 20 nm (red dash dotted), 25 nm (green dotted) and 30 nm
(black short dash dotted).}}
\end{center}
\end{figure}

Notice that the AlGaAs/GaAs heterostructure has $\varepsilon_r \approx 1$, i.e. the dielectric mismatch in this case is negligible. In fact, one can verify in Fig. \ref{fig1}(a) that for AlGaAs/GaAs, $\Delta E_R \approx 0$ for any value of wire radius. AlGaAs/GaAs core-shell wires have been widely experimentally studied for many reasons, e.g. the possibility of obtaining lattice matched defect-free samples, where the AlGaAs shell reduces the surface-related nonradiative recombination, by passivating the GaAs core. \cite{Kang} However, the results in Fig. 1(a) suggest that this heterostructure is actually not the best choice for investigating dielectric mismatch effects. As we will demonstrate onwards, the case of GaN/HfO$_2$ heterostructures is particularly interesting. 

The growth of nitride-based cylindrical core wires, \cite{Jiang} as well as HfO$_2$ shells, \cite{zhu} has been reported recently, which makes us believe that such a structure could be experimentally realized in near future. Figure \ref{fig1}(b) shows the ground state energies $E_{0}$ (i.e., for $n = 1$ and $l = 0$) for each carrier in a core-shell
GaN/HfO$_2$ heterostructure as a function of the core radius. As
usual, as the radius increases, the carriers confinement becomes
weaker, leading to a reduction in the ground state energy, which
is clearly observed for the electron and the light hole in this
case. Conversely, the heavy hole energy shows a non-monotonic
dependence on the radius, namely, it decreases for smaller radii
but starts to increase as the radius becomes larger, suggesting a
different confinement regime for this carrier. In fact, such a
behavior is analogous to the one reported in the literature for
GaN/HfO$_2$ quantum wells. \cite{pereira2} In the quantum wells case, such
a non-monotonic behavior of the confinement energy was
demonstrated to be a consequence of the fact that $\epsilon_r < 1$
in this heterostructure, leading to an attractive potential which
pulls the carrier towards the interface between the materials.
Analogously, this suggests the existence of an interfacial
confinement of the heavy hole in the NW. For smaller radii, the
quantum confinement due to the bands mismatch is still dominant,
but as the radius is enlarged, the energy decreases and eventually
enters the negative energy domain, where the $hh$ becomes bound to
the core-shell interface due to the self-image potential. Being
compressed towards the interface, the $hh$ experiences an increase
in its energy as the radius is further enlarged.

Notice that the sign of the confinement energy indicates whether a charge carrier is confined in the core or at the
interfacial region. Without dielectric mismatch, or even for
$\epsilon_r > 1$, the confinement energy is always positive, since
the bottom of the confinement potential $V_T^i$ is always either
zero or larger. Conversely, the presence of a $\epsilon_r < 1$ mismatch is
responsible for negative cusps on the potential at the interfacial
region. \cite{pereira,slachmuylders} Hence, a negative confinement energy suggests an interfacial confinement. In fact, the heavy hole is the only carrier with negative energy in Fig. 1 (b) and it is indeed the only one exhibiting non-monotonic behavior of the confinement energy as the radius increases. 

The consequences of the interfacial confinement are numerous. For
instance, for some combinations of core-shell materials, one might
find holes in the interface and electrons in the core. Such an
electron-hole spatial separation leads to a reduction of the
overlap between their wavefunctions, reducing the oscillator
strength and, consequently, the recombination rate. The overlaps between electron and light ($lh$, top) and heavy ($hh$, bottom) hole wave functions are shown as a function of the core wire radius $R$ in Fig. 2 (a), for the same combinations of core and shell materials in Fig. 1 (a). For all the heterostructures shown, the overlap value is slightly lower than unit for small radii. This is just a consequence of the fact that wave functions for different charge carriers penetrate with different depths into the barriers, and this effect is more pronounced when the NW radius is small. In the AlGaAs/GaAs (blue), GaAs/ZnSe (green), and Si/SiO$_2$ (black) cases, where the dielectric constants ratios are $\epsilon_r > 1$, the overlaps for both $el-lh$ and $el-hh$ pairs simply converge to 1 as the radius increases, which means that electrons and holes are practically equally distributed in space for larger wire radius. However, the overlaps in the $\epsilon_r < 1$ heterostructures exhibit a maximum at moderate $R$ (around 10-20 nm), and decrease after this value, indicating different distributions of electron and holes functions for large $R$. Although this effect is also present for the Si/HfO$_2$ heterostructure, it is much stronger in the GaN/HfO$_2$ case (red), specially for the $el-hh$ pair.

In order to help us to understand the heavy hole interfacial confinement and the consequently different spatial distribution of charge carriers, Fig. \ref{fig2} (b) illustrates the the total confinement potential (black solid), along with the normalized ground state wave functions, for electrons (top) and heavy holes (bottom) in a $R = 30$ nm GaN/HfO$_2$ core-shell NW. The electron wave function considering dielectric mismatch effects (black dashed) is still confined mostly in the vicinity of the central axis of the core, even at such a large radius, and its width is wider than that without dielectric mismatch (blue dash dotted). On the other hand, the heavy hole wave functions, depicted in the bottom panel of Fig. \ref{fig2} (b) for different values of the core wire radius $R$, are clearly confined at the interface, due to the cusp formed in this region by the self-energy potential (see black solid line). Notice that as the radius increases, the $hh$ wave function is squeezed towards the interface, explaining the increase of the energy in the heavy hole $E_0$ curve in Fig. 1 (b). Due to its larger effective mass, as compared to that of the other charge carriers, the heavy hole is indeed expected to be more strongly confined in the interfacial cusps. Moreover, heavy holes in GaN/HfO$_2$ also have larger effective mass as compared to any charge carrier in Si/HfO$_2$, explaining why the interfacial confinement effect is weaker in the latter case. 

\begin{figure}[t]
\begin{center}
\includegraphics[width= 1.0\linewidth]{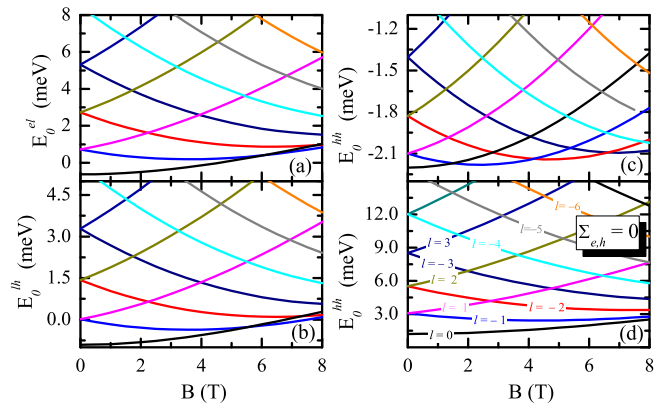}
\caption{\small{\label{fig3} Confinement energies as a function of
the magnetic field for (a) electrons, (b) light and (c) heavy
holes, considering $n = 1$ and different values of angular
momentum index $l$, in a $R$ = 30 nm GaN/HfO$_2$ NW. (d) The results for
a heavy hole in the same system, but neglecting image charge
effects.}}
\end{center}
\end{figure}

Another interesting effect arising from the interfacial confinement in quantum wires comes from their topology: systems where the charge carriers are confined around a core are known to produce an interesting effect when a magnetic field is
applied perpendicular to the confinement plane, namely, angular momentum
transitions occur, even for the ground state, as the magnetic
field intensity increases, which is reminiscent of the
Aharonov-Bohm (AB) effect. \cite{silva} In order to illustrate
this, the electron and hole energy behaviors under such a magnetic field
were numerically \cite{silva} calculated and shown in Fig. \ref{fig3} as a function of
the field amplitude $B$ for the GaN/HfO$_2$ NW. All the carriers have $E_0 < 0$, suggesting that their ground states are all interfacially confined. A large ($R = 30$ nm) radius is considered, in order to enhance the interfacial confinement effect (see Fig. 2 (b)) and reduce the AB period, since this period is inversely proportional to the
average radius of the carrier wave function. Indeed, the ground state energy exhibits AB oscillations for all carriers
in this case, although they are much more evident for the $hh$. Due to their considerably smaller effective masses, \cite{pereira3} $el$ and $lh$ are just weakly confined in the interface, and still
have a large wave function tail spreading inside the core, explaining
the large AB period. The first ground state transition for these carriers is not so visible in Figs. \ref{fig3} (a, b), and occurs at $B \approx 6$ T. On the other hand, for the $hh$, the AB transitions are quite clear, occurring at smaller magnetic fields, with a period $B \approx 2.5$ T. One easily verifies that the
results neglecting the self-energy term in Fig. \ref{fig3} (d) are qualitatively different, showing a
more conventional behavior with the magnetic field for a core-shell NW, namely, without angular momentum transitions for
the ground state.

The interfacial confinement is also expected to have an important
role on the electrons mobility along the core axis, since this
property depends on the electrons density in the central region of
the wire, which is suppressed in the case of interfacial
confinement. However, a more detailed investigation on this issue
is left for future works.

In summary, we have theoretically investigated the charge carriers
confinement in core-shell nanowires with strong dielectric
mismatch. Our results predict that, for specific configurations of
the system, the carriers may be confined at the core-shell
interface. Such interfacial confinement leads to drastic
modifications on the electronic properties of the NW,
specially under an applied magnetic field, where angular momentum
transitions occur for the ground state, due to the AB effect. A decrease in the oscillator strength of the electron-hole pairs in $\epsilon_r < 1$ core-shell quantum wires is also predicted for larger wire radii, which directly affects their recombination rates. We believe that our results will spur on future experimental investigations on core-shell wires made out of high-k materials, contributing for a better understanding of these systems.

This work has received financial support from the Brazilian
National Research Council (CNPq), CAPES, PRONEX/CNPq/FUNCAP and
FAPEMAT (Process No. 39788/2009).


\begin{thebibliography}{99}
\bibitem{mohan}P. Mohan, J. Motohisa, and T. Fukui, Nanotechnology, \textbf{16}, 2903 (2005).
\bibitem{persson}A. I. Persson, M. W. Larsson, S. Stenstrom, B. J. Ohlsson, L. Samuelson, L. R. Wallenberg, Nano Materials \textbf{3} 677 (2004).
\bibitem{chen} C. Chen, S. Shehata, C. Fradin, R. LaPierre, C. Couteau, and G. Weihs, Nano Lett. \textbf{7}, 2584 (2007).
\bibitem{tsivion}D. Tsivion, M. Schvartzman, R. P. Biro, P. Huth, and E. Joselevich, Science \textbf{333}, 1003 (2011).
\bibitem{Jena} D. Jena and A. Konar, Phys. Rev. Lett. \textbf{98}, 136805 (2007).
\bibitem{zhu} X. Zhu, Q. Li, D. E. Ioannou, D. Gu, J. E. Bonevich, H. Baumgart, J. S. Suehle, and C. A. Richter, Nanotechnology \textbf{22}, 254020  (2011).
\bibitem{Jiang}X. Jiang, Q. Xiong, S. Nam, F. Qian, Y. Li, and C. M. Lieber, Nanoletters \textbf{7} 3214 (2007).
\bibitem{li}B. Li, B. Partoens, F. M. Peeters, and W. Magnus Phys. Rev. B \textbf{79}, 085306 (2009).
\bibitem{silva}J. Costa e Silva, A. Chaves, J. A. K. Freire, V. N. Freire, and G. A. Farias, Phys. Rev. B \textbf{74}, 085317 (2006).
\bibitem{slachmuylders} A. F. Slachmuylders, B. Partoens, W. Manus, and F. M. Peeters, Phys. Rev. B \textbf{74}, 235321 (2006).
\bibitem{Supplementary} For more details about our calculations, see the supplementary material.
\bibitem{pereira}T. A. S. Pereira, J. S. de Sousa, J. A. K. Freire, and G. A. Farias, J. Appl. Phys. \textbf{108}, 054311 (2010).
\bibitem{Kang} J.-H. Kang, Q. Gao, H. J. Joyce, H. H. Tan, C. Jagadish, Y. Kim, Y. Guo, H. Xu, J. Zou, M. A. Fickenscher, Leigh M. Smith, Howard E. Jackson, and J. M. Yarrison-Rice, Cryst. Growth Des. 11, 3109 (2011).
\bibitem{pereira2}T. A. S. Pereira, J. S. de Sousa, G. A. Farias, J. A. K. Freire, M. H. Degani, and V. N. Freire, Appl. Phys. Lett. \textbf{87}, 171904 (2005).
\bibitem{pereira3}T. A. S. Pereira, J. A. K. Freire, V. N. Freire, G. A. Farias, L. M. R. Scolfaro, J. R. Leite, and E. F. da Silva, Appl. Phys. Lett. \textbf{88}, 242114 (2006).

\end{thebibliography}
\end{document}


\title {SUPPLEMENTARY MATERIAL}

\maketitle

Our system consists of a semiconductor cylindrical nanowire (core region) of radius $R$, surrounded by a different material (shell region). The materials parameters are assumed to change abruptly from the core to the shell regions. Within the adiabatic approach and the effective mass framework, the Hamiltonian for each carrier $i$ = $el$ (electron), $hh$ (heavy hole) or $lh$ (light hole) is given by \cite{silva}

\begin{equation}
\label{eq1} H_i = \left(\bold{p}_i-q\bold{A}\right)\frac{1}{2m_i}\left(\bold{p}_i-q\bold{A}\right) + V_T^i\left(\rho_i \right),
\end{equation}

\hfill
\hfill

\noindent where $\bold{p}_i \left(=-i \hbar \nabla \right)$ is the momentum operator and $m_i$ is the effective mass for the $i$-th carrier, which depends on the radial coordinate, due to the different masses in the core and shell materials. The wire is assumed to be infinitely long in the $z$-direction, allowing us to use a plane wave solution $\exp(ik_zz_i)$ for this direction, leading to an energy $E_z = \hbar^2k_z^2\big/2m_i$ for this direction, which is set as the energy referential in our calculations, i.e. $E_z = 0$. Besides, we take advantage of the circular symmetry of our system and choose $\exp(il\theta_i)$ as the solution for the angular coordinate, where $l$ is the angular momentum index. In the presence of a $\vec B = B\hat z$ magnetic field, taking the symmetric gauge $\vec A = B\rho_i\hat \theta_i/2$ for the vector potential, the Hamiltonian eventually assumes the form

\begin{eqnarray} \label{eq.Ham}
\label{eq2} H_i = -\frac{\hbar^2}{2}\left(\frac{1}{\rho_i}\frac{\partial}{\partial \rho_i}\frac{\rho_i}{m_i(\rho_i)}\frac{\partial}{\partial \rho_i}  - \frac{l^2}{m_i(\rho_i)R^2}\right) \nonumber \\ \hspace{3cm} -i\hbar\omega_c\frac{l}{2} + \frac{1}{8}m_i\omega_c^2\rho_i^2 + V_T^i(\rho_i).
\end{eqnarray}

\hfill
\hfill

\noindent The total confinement potential $V_T^i(\rho_i) = \Delta E_i(\rho_i)+\Sigma_i(\rho_i)$ is given by the sum of band edges discontinuities $\Delta E_i(\rho_i)$ and the self-energy potential $\Sigma_i(\rho_i)$. The latter term appears due to the dielectric mismatch, and its calculation is based on the method of the image charges, similarly to Refs. \onlinecite{slachmuylders, banyai}. The self-energy potential $\Sigma_i(\rho_i)$ inside the core region, due to a carrier located in the core region, is calculated by solving the Poisson equation

\begin{equation}
\label{eq3}\varepsilon_1 \nabla ^2 V\left(r,r' \right) =  - e\delta \left( r - r' \right),
\end{equation}

\hfill
\hfill

\noindent where $\varepsilon_1$ is the dielectric constant of the core region, and the solution $V_{in} \left(r,r' \right)$ in this region is given by

\[
V_{in}\left(r,r' \right) = \frac{e}{4\pi \varepsilon_1}\left[ \frac{1}{\left|r - r' \right|} + \frac{2}{\pi}\left( \varepsilon_r - 1 \right)\sum\limits_{m = - \infty }^{m =  + \infty } {e^{im\left(\theta  - \theta' \right)}}\right.
\]
\begin{equation}
\label{eq4} \left. \times \int\limits_0^\infty  {dk\cos \left[ {k\left( z - z' \right)} \right]} {C_m}\left( kR,\varepsilon_r \right){I_m}\left( {k\rho } \right){I_m}\left( {k\rho '} \right) \right],
\end{equation}

\hfill
\hfill

\noindent with

\begin{equation}
\label{eq5} C_m\left(kR,\varepsilon_r \right) = \frac{K_m \left(kR \right)K'_m \left(kR \right)}{I_m\left(kR \right)K'_m\left(kR \right) - \varepsilon _r I'_m\left(kR \right)K_m\left(kR \right)},
\end{equation}

\hfill
\hfill

\noindent where $I_m(x)$ ($K_m(x)$) is the $m$-th order modified Bessel function of the first (second) kind and $\varepsilon_r = \varepsilon_1/\varepsilon_2$. 

When the carrier is located in the shell, the Poisson equation to be solved in this region is

\begin{equation}
\label{eq6}\varepsilon_2 \nabla ^2 V\left(r,r' \right) =  - e\delta \left( r - r' \right),
\end{equation}

\hfill
\hfill

\noindent where $\varepsilon_2$ is the dielectric constant of the shell region, and the solution $V_{out} \left(r,r' \right)$ for this equation is given by

\[
V_{out}\left(r,r'\right) = \frac{e}{4 \pi \varepsilon_2}\left[\frac{1}{\left|r-r' \right|} + \frac{2}{\pi}\left( \varepsilon_r - 1 \right)\sum\limits_{m = - \infty }^{ + \infty }{e^{im(\theta  - \theta ')}}  \right.
\]
\begin{equation}
\label{eq7}
\left.  \times \int\limits_0^\infty {dk\cos [k(z - z')]D_m\left(kR,\varepsilon_r \right)K_m(k\rho )K_m(k\rho ')} \right]
\end{equation}

\hfill
\hfill

\noindent with

\begin{equation}
\label{eq8}
D_m\left(kR,\varepsilon_r\right) = \frac{I_m(kR)I'_m(kR)}{I_m(kR)K'_m(kR) - \varepsilon_rI'_m(kR)K_m(kR)}
\end{equation}

\hfill
\hfill

The self-energy potential comes from all the interactions between the carrier and its own image charges, which in the core region is calculated by solving the integral

\begin{equation}
\label{eq9}\Sigma_i^{in}\left(\rho_i \right) = \frac{e}{2} \int {  \delta\left(r-r_i \right) V_{in}\left(r,r_i\right)dr},
\end{equation}

\hfill
\hfill

\noindent that results on 

\begin{equation}
\label{eq10}\Sigma_i^{in}\left(\rho_i\right) = \frac{{{e^2}}}{{4{\pi ^2}{\varepsilon _1}}}\left( {\varepsilon_r - 1} \right)\sum\limits_{m =  - \infty }^{ + \infty } {\int\limits_0^\infty  {dk{C_m}\left( {kR,\varepsilon_r} \right)I_m^2(k\rho_i)}}
\end{equation}

\hfill
\hfill

\noindent for electrons and holes ($i=el, lh, hh$) in this region. In the shell region, the self-energy potential comes from the integral

\begin{equation}
\label{eq11}\Sigma_i^{out}\left(\rho_i \right) = \frac{e}{2} \int {  \delta\left(r-r_i \right) V_{out}\left(r,r_i\right)dr},
\end{equation}

\hfill
\hfill

\noindent that results on

\begin{equation}
\label{eq12}\Sigma_i^{out}(\rho_i) = \frac{{{e^2}}}{{4{\pi ^2}{\varepsilon _2}}}\left( {\varepsilon_r - 1} \right)\sum\limits_{m =  - \infty }^{ + \infty } {\int\limits_0^\infty  {dk{D_m}\left( {kR,\varepsilon_r} \right)}}K_m^2(k\rho_i)
\end{equation}

\hfill
\hfill

Notice the solution in the first case can be compared to the second case just by changing the modified Bessel functions of the first type, $I_m^2(k\rho_i)$  by those of the second type, $K_m^2(k\rho_i)$. The self-energy potentials defined by $\Sigma_i^{in}(\rho_i)$  in Eq. (\ref{eq10}) and $\Sigma_i^{out}(\rho_i)$ in Eq. (\ref{eq12}) are eventually obtained by numerically solving the integrals, and add on $V_T^i(\rho_i)$. Finally, the Schr\"odinger equation for the Hamiltonian Eq. (\ref{eq.Ham}) is discretized in a finite differences scheme, \cite{Peeters} and the resulting matrix eigenvalues problem is numerically solved by standard computational routines, in order to obtain the energy spectrum of the quantum wire system.